# Thermal transport properties and some hydrodynamic-like behavior in 3D topological semimetal ZrTe$_5$

Chang-woo Cho[1,2], Peipei Wang[1], Fangdong Tang[1], Sungkyun Park[2], Mingquan He[3], Rolf Lortz[4], Genda Gu[5], Qiang Li[5,6], and Liyuan Zhang[1,+]

[1]Department of Physics, Southern University of Science and Technology, Shenzhen 518055, China

[2]Department of Physics, Pusan National University, Busan 46241, South Korea

[3]Low Temperature Physics Lab, College of Physics & Center of Quantum Materials and Devices, Chongqing University, Chongqing 401331, China

[4]Department of Physics, The Hong Kong University of Science and Technology, Clear Water Bay, Kowloon, Hong Kong

[5]Condensed Matter Physics and Materials Science Department, Brookhaven National Laboratory, Upton, NY 11973, USA

[6]Department of Physics and Astronomy, Stony Brook University, Stony Brook, New York 11794-3800, USA

## Abstract

**Hydrodynamic fluidity in condensed matter physics has been experimentally demonstrated only in a limited number of compounds due to the stringent conditions that must be met. Herein, we performed thermal and electrical transport experiments in three-dimensional topological semimetal ZrTe5. By measuring the thermal properties in a wide temperature range, two representative experimental evidences of the hydrodynamics are observed in temperature window between the ballistic and diffusive regimes: a faster evolution of the thermal conductivity than in the ballistic regime and the non-monotonic temperature-dependent effective quasiparticle mean-free-path. In addition, magneto-thermal conductivity results indicate that charged quasiparticles, as well as phonons, may also play an important role in this hydrodynamic-like flow in ZrTe5.**

[+] corresponding author: zhangly@sustech.edu.cn

# Introduction

In insulators, heat is mainly carried by phonons. This phonon-dominant heat conduction is described by Fourier's law, in which phonons scatter from other phonons, impurities, and boundaries [1-3]. This process takes place through the momentum-relaxing process known as Umklapp scattering (U-scattering). During this process, heat fluxes are dissipated and the crystal momentum is not conserved [1-3]. On the other hand, at a sufficiently low temperature $T$, Fourier's law no longer holds, where the crystal momentum is conserved thanks to the dominant Normal scattering (N-scattering) [4-6]. These two types of scattering mechanisms are known for a diffusive and a ballistic regime, respectively, and have been widely studied in many solids [7-11].

Meanwhile, Gurzhi proposed a viscous flow driven by the heat carriers when N-scattering is abundant in the overlapping two regimes [12]. Since then, it has been called hydrodynamic flow due to its analogy with macroscopic transport phenomena in water fluids [13]. When phonons represent the primary heat carriers in solids, two significant characteristics are known as the Poiseuille flow and the second-sound wave [6,14]. The former is characterized by a steady-state phonon flow in which thermal resistance diffuses due to the boundary scattering combined with N-scattering [15,16]. In comparison, the latter involves wave-propagation of a $T$-gradient without significant attenuation [6,17,18].

Despite the fascination of phonon-hydrodynamics (PH) in solid state systems, experimental observation is rare. Moreover, it is found only in a narrow $T$-window at a remarkably low $T$, where abundant N-scattering and a suitable sample size are additionally required. For instance, the reported $T$-window of Poiseuille flow in suspended graphene was only 0.5 K at about 1 K. [19]. One reason for this practical difficulty is that U-scattering overwhelms N-scattering in almost every $T$-range except at significantly low $T$. For these reasons, PH behavior has been experimentally confirmed in only a handful of compounds, such as solid He-3 [20] and He-4 [21], Bi [22], black P [16], $SrTiO_3$ [23], and graphite [24,25]. Therefore, the search for new materials in which hydrodynamics contributed through phonons or other collective excitations is of great interest to the condensed matter community.

In this study, we performed thermal and electrical transport experiments for topological semimetal $ZrTe_5$ to investigate the hydrodynamic property. In fact, the $ZrTe_5$ study was initiated decades ago due to its considerable thermoelectric performance and resistivity anomaly [26,27]. Recently, it has gained renewed attention due to non-trivial topological phenomena such as a 3D quantum Hall effect [28], a quantum spin Hall effect on a monolayer [29], and a chiral magnetic effect [30]. Moreover, it has been reported that bulk $ZrTe_5$ sits at the boundary between a weak- and a strong-topological phase, so that an external perturbation easily affects its topology [31-33]. Herein, we present experimental evidence for PH by observing a faster evolution of the thermal conductivity $\kappa$ than in the ballistic regime. In contrast to the conventional PH, we find an unexpected thermal transport behavior in a hydrodynamic regime, which could be attributed to the charged quasiparticles. After reviewing several scenarios, we suggest that hydrodynamic flow is mainly led by phonons, but presumably weak coupled to charged quasiparticles. Our findings have important implications for ongoing research on the various possible types of hydrodynamics, especially in a three-dimensional topological semimetal.

## Methods

In the experimental setup, we used ultrahigh quality ZrTe$_5$ single crystals, grown by the tellurium flux method. Details of the sample growth and structural properties can be found elsewhere [28,34,35]. Thanks to the relatively large size of the single acicular crystals (length, *l* x width, *w* x thickness, *t*; Sample #1: 3.0 x 0.4 x 0.3 mm$^3$, Sample #2: 3.2 x 0.3 x 0.1 mm$^3$, Sample #3: 2.9 x 0.3 x 0.2 mm$^3$), we were able to perform the electrical and thermal transport experiments on the same bulk samples. In the main text, we defined the longest (shortest) dimension as along the *a*-axis (*b*-axis), corresponding to the ZrTe$_3$ chain (stacking layer) direction.

In the transport experiments, we performed the electrical resistivity measurements by the standard Hall bar method, using an alternating current with an amplitude of 0.01-0.1 *m*A and a frequency of 10-20 Hz. The magnetic field *B* was applied in the perpendicular direction to the *ac*-plane. In order to measure the thermal transport of such a needle-shaped ZrTe$_5$ crystal, we used a well-known steady-state method with one-heater and three-thermometers as shown in **Fig. 1a-c**. One end of a ~3.0 mm long sample was attached to a copper heat sink, while a small chip-like 100 Ohms resistor and three well-calibrated Cernox thermometers were suspended from glass fibers. To minimize heat loss, thin Pt/W wires (25 *um*) were used between all electrical devices and electrodes on the holder, while thick Ag wires (100 *um*) were connected to the sample for the best thermal equilibrium state during the measurement. To eliminate spurious longitudinal (or transverse) components, we measured and averaged every transport experiment in opposite *B*-field directions. For obtaining more accurate thermal transport quantities, we also measured the thermal conductivity of high-purity brass under the identical experimental environment and calibrated it accordingly (see Supplementary **Fig. S1**). Since the sensitivity of the thermometers as a thermal detector becomes weaker towards higher *T*, we switched to the thermocouple method to record the thermal gradient in the high *T* regime (*T* > ~20 K). In the overlapping range (about 10-20 K), we confirmed the consistent $\kappa$ results within the error bars; an example for Sample #3 is presented in supplementary **Fig. S2**.

## Results

In the following, we will examine the first evidence of hydrodynamic flow. In a PH regime, $\kappa$ should evolve faster than a *T*$^3$-dependence. To test this, we plot the *T*-dependent total thermal conductivity $\kappa_{tot}$ in the ZrTe$_5$ single crystals, which are shown in **Fig. 2a**. The electronic contribution of thermal conductivity $\kappa_{WF}$ (solid lines) is presented in **Fig. 2b**. Note that only $\kappa_{tot}$ is a directly measured value, whereas $\kappa_{WF}$ is extracted from the Wiedemann-Franz (WF) law ($\kappa_{WF} = \frac{T}{\rho}L_0$, Lorenz number $L_0 = 2.44 \times 10^{-8}\ W\Omega K^{-2}$) based on our electrical resistivity data. For clarity, it is plotted on a log-log scale here. In the high $T$-regime (~200 - 300 K in Sample #1, and ~30 - 300 K in Sample #2 and #3), it follows nearly $1/T$ dependence in all the samples, meaning that the U-scattering is the most prominent process in this range. After passing through the $\kappa_{tot}$ peak, it starts to decrease, indicating the N-scattering process begins to dominate. In the case of the thickest sample (#1, $t = 0.3$ mm), no hydrodynamic features are seen, whereas Sample #2 and #3 deviate from the line proportional to $T^3$ (dashed line in **Fig. 2a**) in a low $T$ regime. For Sample #3 ($t = 0.2$ mm), it first shows a downward kink-like anomaly, presumably considered phonon-drag effect, just below the $T$ where the maximum occurs. With further cooling, the slope of $\kappa_{tot}$ gradually increases towards low $T$ and exceeds a $T^3$-dependence in a range of 0.8 to 2.0 K. This is clearer with a wide $T$-window in the thinnest sample (#2, $t = 0.1$ mm). By comparison $\kappa_{tot}$ and $\kappa_{WF}$, it should be mentioned that, if $\kappa_{tot}$ assumes the summation of phononic and electronic contribution, the phonon thermal conductivity $\kappa_{ph}$ dominates by more than one order of magnitude across the entire $T$-range (see **Fig. 2a** and **b**). Another thing is that the $\kappa_{tot}$ must converge to $\kappa_{WF}$ at sufficiently low $T$ because the thermal energy at low $T$ is mainly transferred from the charge carriers. However, we see no convergence up to the experimental low $T$-limit of 0.7 K. This is due to the comparatively high-purity crystallinity and extremely low carrier density at low $T$ (see **Fig. S3b and c**), so that phonons still play a crucial role at low $T$. We thus deduce that such a hydrodynamic feature is attributed to predominantly phonons.

Next, we examine the $B$-field dependence of thermal transport. In **Fig. 3a**, we present the longitudinal thermal conductivity $\kappa_{xx}$ as a function of $B$-field in Sample #3, which measured at $T = 0.81$ K. For comparison with the electronic contribution, we plot together with $\kappa_{WF}$ (solid red line in **Fig. 3a**) measured at $T = 0.70$ K. Two things are worth noting here. First, one sees an apparent thermal quantum oscillation that is in complete agreement with the electronic quantum oscillations. Although phonons still play a dominant role up to our experimental low-$T$ limit of 0.7 K, it means that the contribution of charged particles among thermal carriers increases when $T$ is lower. Second, it hardly responds to $\kappa_{xx}$ when the external $B$-fields are sufficiently high. When the quantum oscillations terminate at ~1.5 T, $\kappa_{xx}$ is nearly constant above this field. We also confirm this for Sample #2 that $\kappa_{xx}$ is barely influenced by the $B$-fields regardless of base $T$ (see **Fig. S4**).

An unexpected thing is seen in the $T$-dependent electronic thermal contribution. In general, it is known that $\kappa_{ph}$ does not seriously change by the external $B$-fields, thus we can extract the thermal contribution of charged quasiparticles by subtracting the $\kappa_{xx}(B)$ from the $\kappa_{xx}(0T)$. To do this, we define $\Delta\kappa = \kappa_{xx}(0T) - \kappa_{xx}(B)$ and plotted in **Fig. 3b**, where the magnitude of $B$-field was chosen to be 2.4 T and 5.0 T for Sample #3 and #2, respectively. Surprisingly, this quantity exhibits a distinct deviation in the hydrodynamic window we observed.

To gain a deeper understanding, we carry out the thermal Hall experiment, as this could be a direct probe to study quasiparticle dynamics. **Figure 4a** shows the *B*-field dependence of the thermal Hall resistivity $\omega_{xy}$ $(= \frac{wt}{l}\left(\frac{\Delta T_{xy}}{P}\right)$, where $\Delta T_{xy}$ and *P* denote the *T*-gradient between two points along the transverse direction and the heating power, respectively) in a narrow *B*-field range from -1 to 1 T. For a higher resolution, we recorded the data this time with a continuous field sweep mode. In the main text, only the case of Sample #2 is shown (Sample #3 data are included in supplementary **Fig. S5**). The $\omega_{xy}$ is negligibly small in almost all *B*-fields except for in a very weak field region ($|B|$ < 0.1 T). It is noted that purely phononic thermal contribution cannot be detected by the thermal Hall signal due to its charge neutrality. Thus, zero thermal Hall voltage is acceptable because phonons are the primary heat carriers in ZrTe$_5$ single crystals in this study. Then, the transverse thermal gradient should not be generated under the *B*-fields. Interestingly, an asymmetric thermal Hall feature is found in a weak field region, it becomes stronger as *T* decreases.

The degree of heat deviation can be determined from the thermal Hall angle $\tan\theta_H$. In **Fig. 4b**, $\tan\theta_H$ $(= \frac{\kappa_{xy}}{\kappa_{xx}})$ is plotted as a function of *B*-field with various *T*. The trend is not different from $\omega_{xy}$ versus *B*. It exhibits a significant deviation when the *B*-field is applied near zero-field and is abruptly faded in the region of higher *B*-fields. In **Fig. 4c**, we represent the zero-field-limit ($B \to 0$) of $\tan\theta_H /B$ (hereafter $[\tan\theta_H /B]_0$), which is proportional to the effective mean-free-path of the quasiparticles $l_{QP}$ [36]. The magnitude of $l_{QP}$ can be estimated through the equation $l_{QP} = \frac{\hbar k_F}{e}\frac{\tan\theta_H}{B}$, where $\hbar$ is the Planck constant, $k_F$ is the Fermi wave number, and $e$ is the electron charge [37]. Using the estimation of $k_F \approx 4 \times 10^{-3} \text{Å}^{-1}$ in the *ac*-plane [28], we obtain that the $l_{QP}$ is about 40 *um* at 1.0 K in both samples (#2 and #3), which is notably longer than those previously reported [38-40]. This consequence also supports our extremely clean ZrTe$_5$ samples, so that quasiparticles travel without significant momentum loss. In particular, the $l_{QP}$ at about 0.7 K is by a factor of 5 longer than that of 1.0 K, where its length scale is exceeding to our thinnest sample thickness (#2, *t* = 0.1 *mm*). Another striking feature of $[\tan\theta_H /B]_0$ is the presence of a local minimum (vertical arrows in **Fig. 4c**) corresponding to *T* at ~1.8 K (Sample #2) and ~2.2 K (Sample #3), which can be an additional signature of hydrodynamic flow. These are also in good agreement with the phonon-dominant hydrodynamic regime we observed.

## Discussion

So far, we have shown the hydrodynamic-like features from the thermal transport experiments. The question that arises from our results is how phonon-dominant hydrodynamics could be realized in the semimetallic ZrTe$_5$ and not in an insulator. In terms of the scattering time scale, the U-scattering time grows exponentially as decreasing *T*, while the N-scattering time is given by a power law *T*-dependence. The boundary scattering time must lie between the two for the realization of hydrodynamic flow. Not only are these conditions hardly satisfied intrinsically, but they are also easily affected by impurities. For this reason, the hydrodynamic regime is extremely fragile and has been found in a limited number of compounds with very narrow $T$-windows, therefore it requires high-purity crystallinity. On the one hand, it is also pointing out that instability of the crystal structure may increase the stability of PH by enhancing N-scattering [41,42]. The materials in which a PH was reported, such as Bi, black P, and $SrTiO_3$, are the supporting examples, because these were not ultra-pure systems like pure silicon. Instead, these materials are in the crystal phase boundary, which makes it easier to be dominant N-scattering in the vicinity of the hydrodynamic window [41,42]. In these senses, $ZrTe_5$ can be a strong candidate for a realization of PH experimentally. Since $ZrTe_5$ has been reported to have an intrinsic unstable crystal structure and topology, its physical properties can be easily tuned by changing the growth environment and other external parameters [28,32,33]. Combined ultra-pure crystallinity in this material, the significant N-scattering in the low $T$ originated from the structural instability makes $ZrTe_5$ a perfectly suitable material for observing phonon-dominant hydrodynamics.

Then it is puzzling what kind of collective quasiparticles induces the hydrodynamic flow in our case. Again, the thermal Hall signal is essentially coming from the electronic contribution, since the neutrally charged quasiparticles are not affected by a magnetic field. Indeed, it was first reported the appearance of a sizable phonon thermal Hall effect in $Tb_3Ga_5O_{12}$ [43]. Soon after, this observation stimulated extended follow-up theoretical and experimental studies to uncover the origin, including phonon-magnon interaction [44], spin-phonon interaction [45,46], Berry curvature [47], skew scattering [48], etc. However, the reported phonon-induced thermal Hall effect is in contradiction to ours. While the magneto-transverse temperature difference showed nearly monotonic increment as $B$-fields increase [43], we find no sizable thermal Hall voltage except for a narrow $B$-field range and low $T$-regimes. Furthermore, the performed studies in recent were focused on mostly magnetic materials because they assumed that it is closely linked to magnetic excitation [36,49,50]. This is not the case in $ZrTe_5$. Hence, it is reasonable to say that the hydrodynamic flow in $ZrTe_5$ is unlikely to be due to a purely phononic attribution. Although we demonstrate that the heat in $ZrTe_5$ is dominantly carried by the phonons, the electron-electron hydrodynamic scenario is still valid. In the results of the zero-field-limit electronic Hall-angle ($[\tan\theta_e/B]_0$, inset of **Fig. 4c**), we can test it. It increases steadily as $T$ drops to ~10 K, and then nearly saturates at low $T$. This indicates that the electron-electron scattering process below 10 K is virtually unaffected by the entire scattering system. From this, we rule out the pure electron-electron fluid scenario.

The next possibility is an electron-phonon fluid in which the electron-phonon scattering process is the fastest, so their momentum can be quasi-conserved. For electron-phonon cases studied previously, the results resembled ours to some extent, since there is a significant violation of $L/L_0$ [51]. However, the sign of $L/L_0$ is at odds with the present results as shown in **Fig. 5**, implying that our system is much closer to a PH-like fluid. Moreover, although we find the signature looking like phonon-drag effect in all samples (vertical arrows in **Fig. 2a**), it is hard to conclude at present if this phonon-drag effect is closely related to phonon-hydrodynamics or not. It is because out of the hydrodynamic window. We propose a continued exploration of the phonon-drag effect in this material.

Dirac fluid may be another candidate. According to the previous work of Crossno et al., in which they reported on a deviation of $\kappa_{WF}$ with largely violated the $L/L_0$ at a charge neutrality point in graphene, and they argued that this is indicative of Dirac fluid [52]. Although seemingly similar to the present results (significant violation of $L/L_0$ and nearly charge-neutrality point), our observations are different in principle. In the case of Crossno et al., the Dirac fluid hydrodynamics occurred in the non-degenerate regime [52], but our ZrTe$_5$ is far away in the degenerate regime. Furthermore, they observed a recovery of $L/L_0$ as one moves away from the neutral point [52], but we see no recovery over the entire $T$-window in Sample #2 and #3 existing the hydrodynamics. Given that none of the scenarios are likely to dominate the hydrodynamics in the present results, therefore we cautiously suggest that predominant PH-like flow that weakly coupled to charged quasiparticles in our three-dimensional topological semimetal ZrTe$_5$ crystals.

## Summary

In summary, we have systematically investigated the thermal and electrical transport properties of bulk ZrTe$_5$ crystals. To date, the main effort of hydrodynamic studies is still needed to find the significant features where either electrons or phonons provide the primary scattering. In addition, all transport regimes - ballistic, hydrodynamic, and diffusive – can coexist and be coupled, making it difficult to distinguish purely quasiparticle hydrodynamic phenomena. Using ultrahigh-purity single crystals of ZrTe$_5$, we have found some PH-like Hallmarks as well as the anomalous flow of charged quasiparticles, which is certainly unexpected. However, the underlying physics of the $B$-field induced oscillation of $\kappa_{xx}$ and the origin of thermal Hall effect remain unknown. These require extended theoretical and experimental work beyond the scope of the present study.

## Acknowledgement

We thank Benjamin Piot and Kitinan Pongsangangan for their enlightening discussions. C.-w. Cho is supported by BK4-program from NRF-Korea. S. Park is supported from NRF-Korea (NRF-2021M3H4A6A02045432). M. He acknowledges the support by National Natural Science Foundation of China (11904040), Chongqing Research Program of Basic Research and Frontier Technology, China (Grant No. cstc2020jcyj-msxmX0263). R. Lortz was supported by grants from the Research Grants Council of the Hong Kong Special Administrative Region, China (GRF-16302018, C6025-19G-A). The work at Brookhaven National Laboratory was supported by the U.S. Department of Energy (DOE) the Office of Basic Energy Sciences, Materials Sciences and Engineering Division under Contract No. DE-SC0012704. The work at

Southern University of Science and Technology acknowledges the support by National Natural Science Foundation of China (118744193), China postdoctoral Science Foundation (2020M672760), Shenzhen Distinguished Young Scholar (RCJC20200714114435105), and Tecent Foundation Xplorer Prize.

**Figures**

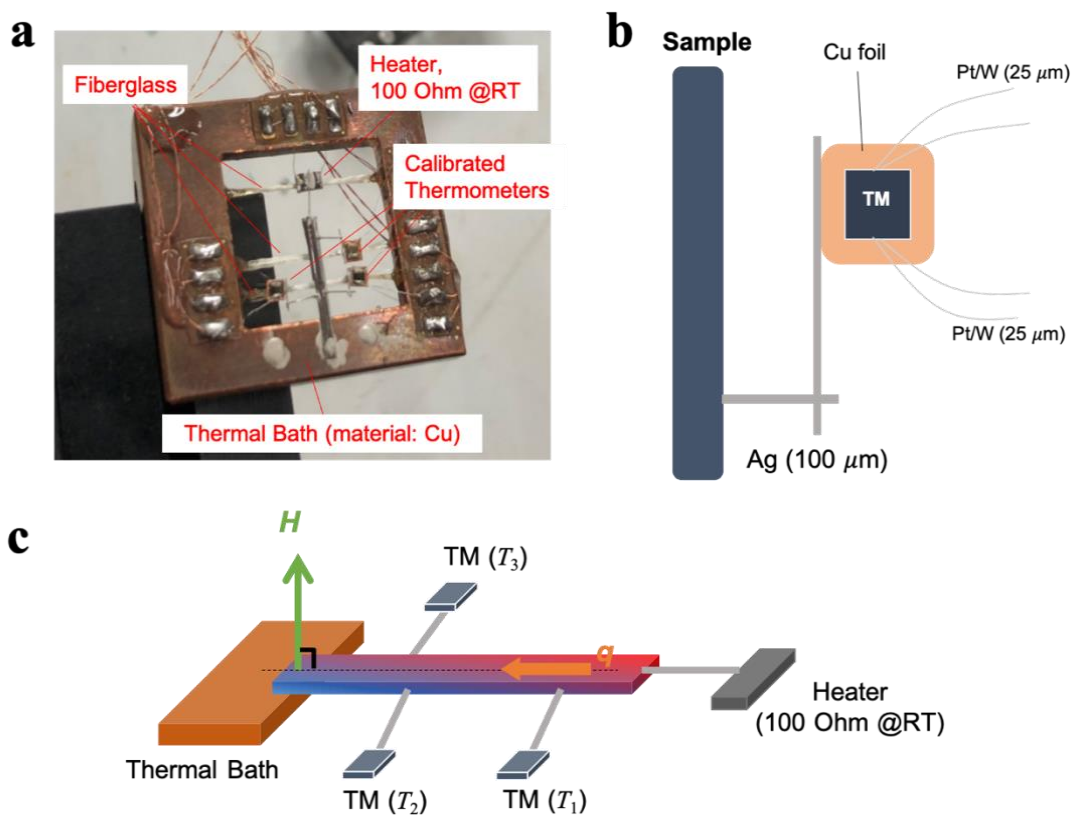

**Figure 1. (a)** Photograph of the thermal conductivity setup used in this study. **(b)** To minimize thermal leakage during the heat flow (from the resistive heater to the thermal bath), we connected the sample to the heater and thermometers through 100 $\mu$m thick Ag-wires. And, the connections for the electrical measurements are made by 25 $\mu$m thin Pt/W-wires since it is a good electrical conductor but a relatively poor thermal conductor. **(c)** Schematic diagram of our thermal conductivity experimental setup.

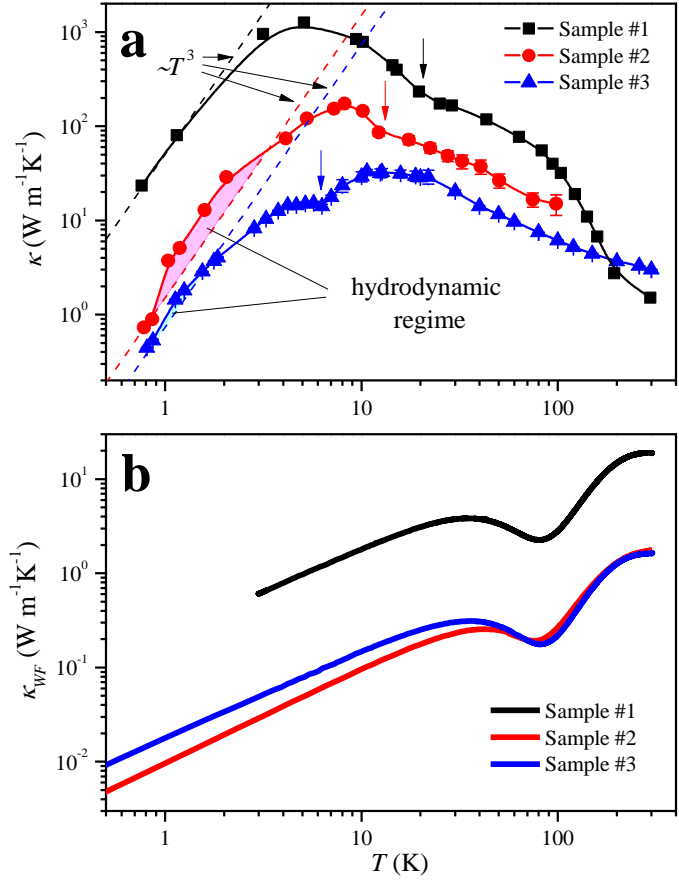

**Figure 2. (a)** Thermal conductivity as a function of temperature in a log-log plot in three different ZrTe$_5$ samples. The squares (Sample #1), circles (Sample #2) and triangles (Sample #3) indicate the total thermal conductivity $\kappa_{tot}$, respectively. All the dash lines are proportional to $T^3$. In Sample #2 and #3, the shaded regions denote the area that exceeds $T^3$-dependence. The vertical arrows denote the temperature that is occurring phonon-drag effect of each sample. **(b)** Temperature-dependent charge carrier thermal conductivity $\kappa_{WF}$, which is calculated according to the Wiedemann-Franz law ($\kappa_{WF} = \frac{T}{\rho}L_0$, where the Lorenz number $L_0 = 2.44 \times 10^{-8}\ W\Omega K^{-2}$). Compared to $\kappa_{tot}$, it is smaller by a factor of hundreds in almost all temperature ranges.

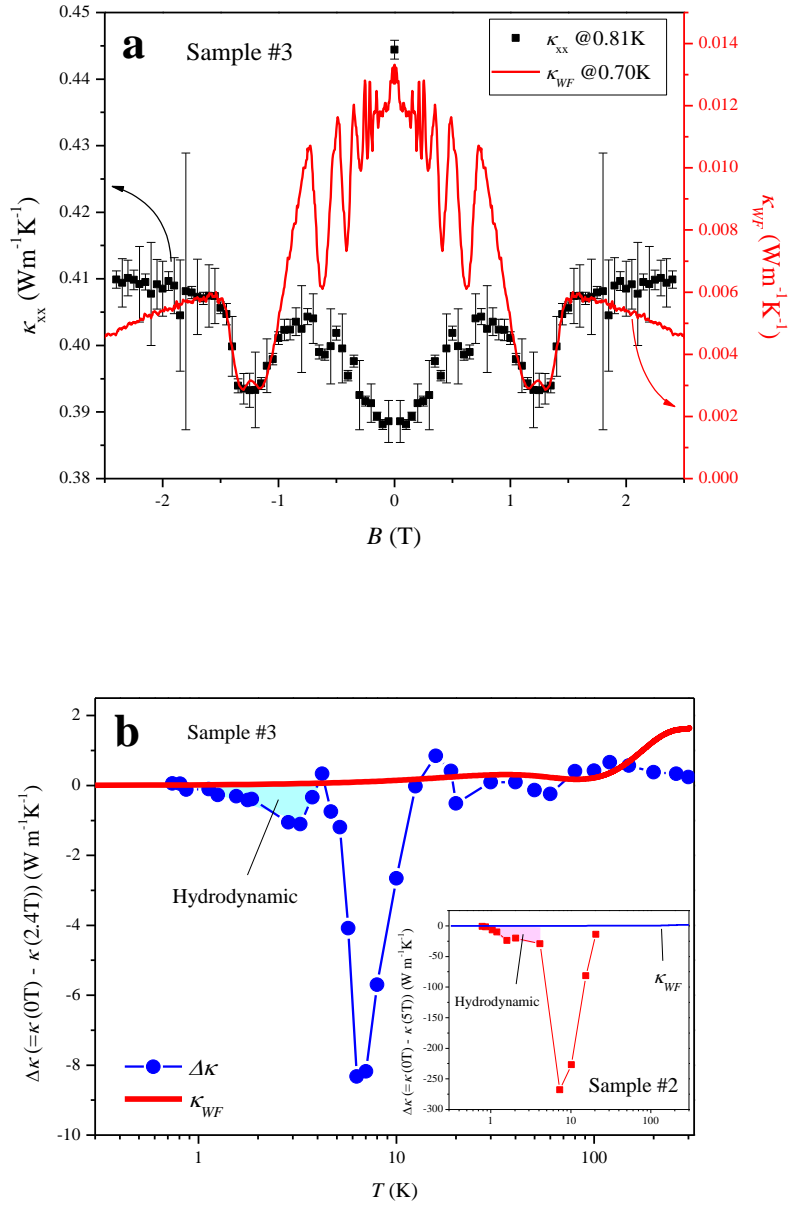

**Figure 3.** **(a)** Longitudinal total thermal conductivity $\kappa_{xx}$ (closed square) and electronically contributed thermal conductivity $\kappa_{WF}$ (red-solid line) as a function of magnetic fields. The data was taken at $T$ = 0.81 K ($\kappa_{xx}$) and $T$ = 0.70 K ($\kappa_{WF}$), respectively. **(b)** Extracted charged quasiparticles contribution to the thermal conductivity of Sample #3 (closed circle). Sample #2 result is added in the inset of **(b)**. Both samples show the deviation of $\Delta\kappa$ in a hydrodynamic regime as shaded. See main text for details.

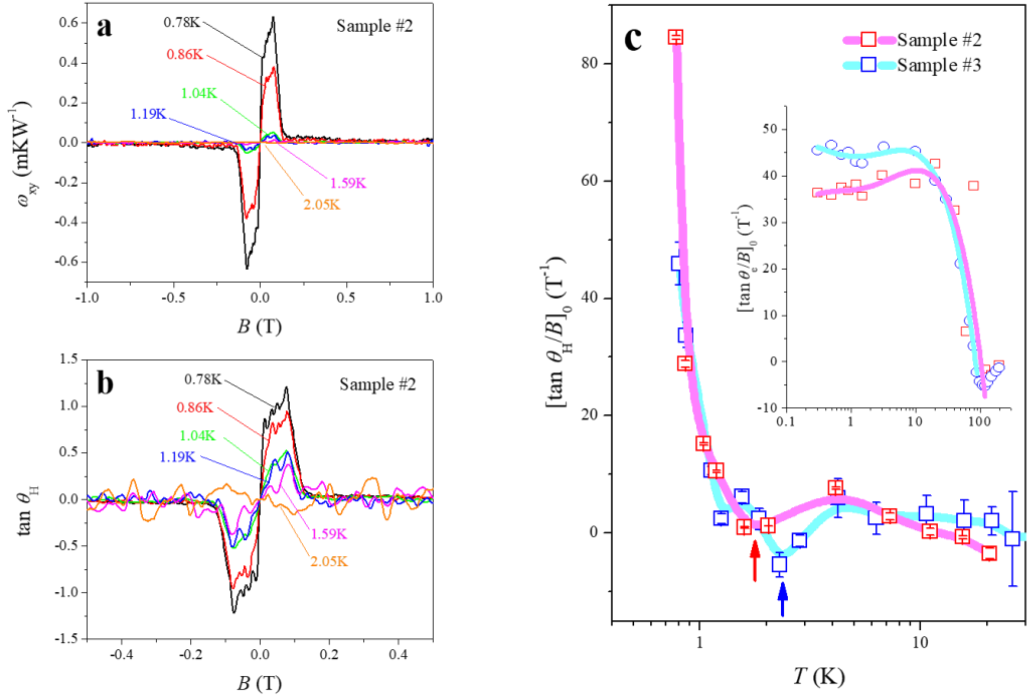

**Figure 4. (a)** Magnetic field dependence of the thermal Hall resistivity $\omega_{xy}$ at various temperatures on Sample #2. **(b)** Tangential Hall angle ($tan\,\theta_H = \frac{\kappa_{xy}}{\kappa_{xx}}$) in a magnetic field range of -0.5 to 0.5 T at different temperatures on Sample #2. **(c)** Temperature-dependent slope of $tan\,\theta_H/B$ in the zero-magnetic-field-limit for Sample #2 and #3. In principle, this quantity is proportional to the mean-free-path of the quasiparticles. The vertical arrows denote the local minima. The inset of **(c)** presents the initial slope of the electronic Hall angle.

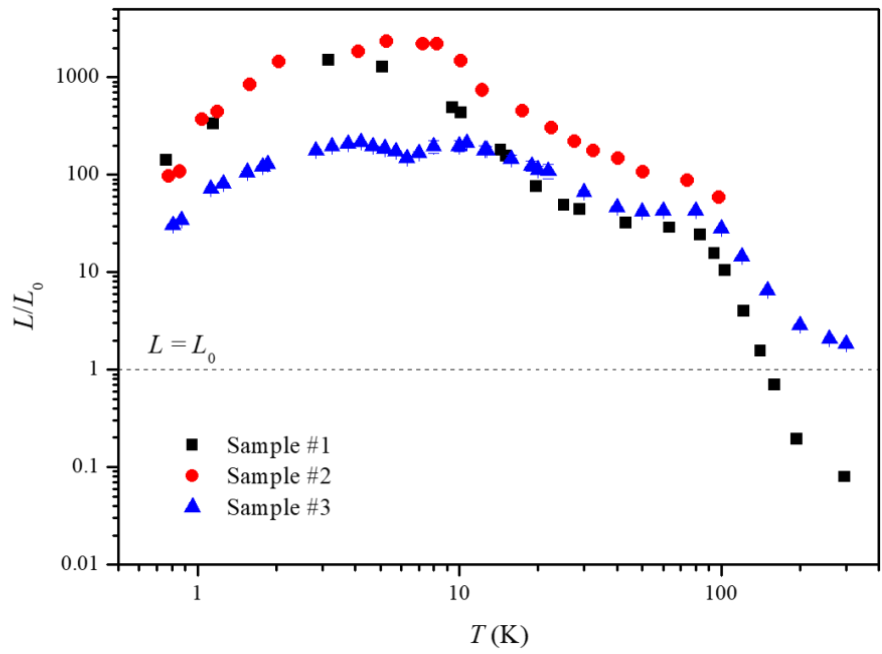

**Figure 5.** Temperature dependence of Lorenz ratio ($L/L_0$) for ZrTe$_5$ single crystals used this study. Horizontal line (dotted) is a guideline when $L=L_0$.